\documentclass[referee]{raa}            % referee version: for submission
\usepackage{graphicx,times}             %for PS/EPS graphics inclusion, new
\input{epsf.sty}                        %for PS/EPS graphics inclusion, old
\input{psfig.sty}                       %for PS/EPS graphics inclusion, old

\def\gm{grand minima}

\def\mm{Maunder minimum}
\def\mc{meridional circulation}
\def\pf{poloidal field}
\def\bl{Babcock--Leighton}
\def\ftdm{flux transport dynamo model}
\def\Rs{R_{\odot}}

\newcommand{\vf}{{\bf v}}

\begin{document}

   \title{Studies of grand minima in sunspot cycles from a flux
	transport solar dynamo model
%\,$^*$
%\footnotetext{$*$ Supported by the National Natural Science Foundation of China.}
}
%   \subtitle{I. Place Your Subtitle Here}

   \volnopage{Vol.0 (200x) No.0, 000--000}      %%preserved for Editor. DOn't remove!
   \setcounter{page}{1}          %%starting page, preserved for Editor. DOn't remove!

   \author{Bidya Binay Karak
%% Please move "\mailto{}" to the corresponding author of the paper.
%% For single author or all the authors from an institute, use "\inst{}" only
%% Here is an example of three authors come from different institutes.
   \and Arnab Rai Choudhuri
     }

   \institute{
            Department of Physics, Indian Institute of Science, Bangalore 560012; {\it
             bidya\_karak@physics.iisc.ernet.in}
      }

\date{Received~~2009 month day; accepted~~2009~~month day}

\abstract{ We propose that grand minima in solar activity are caused by simultaneous fluctuations
in the meridional circulation and the Babcock--Leighton mechanism for the poloidal field generation in the flux transport dynamo model.
We present the following results: (a) fluctuations in the meridional circulation are more effective in producing
grand minima; (b) both sudden and gradual initiations of grand minima are possible; (c) distributions
of durations and waiting times between grand minima seem to be exponential; (d) the coherence
time of the meridional circulation has an effect on the number and the average duration of grand minima, a coherence
time of about 30 years being consistent with observational data. We also study the occurrence of
grand maxima and find that the distributions of durations and waiting times between grand maxima
are also exponential, like the grand minima.  Finally we address the question whether the Babcock--Leighton 
mechanism can be operative during grand minima when there are no sunspots. We show that an
$\alpha$-effect restricted to the upper portions of the convection zone can pull the dynamo
out of the grand minima and can match various observational requirements if the amplitude of
this $\alpha$-effect is suitably fine-tuned. }

\authorrunning{B. B. Karak \& A. R. Choudhuri }            %author_head in even pages
\titlerunning{Grand minima from flux transport dynamo }  % title_head in odd pages

   \maketitle
%% The author head (on even pages) and the title head (on odd pages) will be
%% automatically extracted from \author{} and \title{}. Whenever the title is too long,
%% you will be asked to supply a shorter one by inserting either \authorrunning{} or
%% \titlerunning{} before \maketitle. Anyway, you can specify your own heads.
%%
%%
%% Note: In the following text body of your manuscript, please note several differences from
%%       other major journals:
%% (1) \subsection{Please Capitalize the First Letter of Each Notional Word in Subsection Title}
%% (2) Please Capitalize the First Letter of Each Notional Word in all tables' captions

%
%________________________________________________ sections below
%
\section{Introduction}           %% first-level sections will be auto-capitalized
   
One intriguing aspect of the solar cycle is the occurrence of grand minima when sunspots may not
appear for several decades and a few cycles may go missing. Since the beginning of the telescopic
observations of sunspots, one grand minimum known as the
Maunder minimum occurred during 1645--1715  (Eddy 1976; Ribes \& Nesme-Ribes 1993). 
We have to look for indirect proxy data to infer the occurrences of grand minima at still earlier
times. When the magnetic field of the Sun is weak, 
more cosmic rays reach the Earth's atmosphere, producing larger amounts of the cosmogenic isotopes 
like $^{14}$C and $^{10}$Be. From the study 
of $^{14}$C in old tree rings and $^{10}$Be in polar ice cores, several groups have 
identified a number of grand minima in the past few millennia
(Usoskin et al. 2007; Steinhilber et al. 2012). Particularly, Usoskin et al. (2007) have detected about 27 such events
of low activity in last 11,400~years from the $^{14}$C data. Even when sunspots are not seen,
some of the indirect proxy data indicate the presence of continued oscillations at a
subdued level during grand minima (Fligge et al.\ 1999).
Miyahara et al.\ (2004) found the oscillations to have longer periods during the
Maunder minimum. 
Miyahara et al.\ (2006, 2007) and Nagaya et al. (2012) found this to be true for other
grand minima as well.
 
The aim of the present paper is to investigate whether different aspects of grand minima 
can be explained with a flux transport solar dynamo model. An earlier paper by Choudhuri
\& Karak (2012; hereafter CK12) (also see Karak \& Choudhuri 2013) developed a theoretical model of grand minima by introducing
appropriate fluctuations in our flux transport dynamo model and presented some preliminary
result.  This paper is a continuation of that work and addresses several aspects of the
problem not discussed in CK12.

The \ftdm\ has emerged as the most promising theoretical model for the sunspot cycle in
recent years (Wang et al.\ 1991; Choudhuri et al.\ 1995; Durney 1995; Dikpati \&
Charbonneau 1999; Nandy \& Choudhuri 2002; Charbonneau 2010; Choudhuri 2011; Karak \& Petrovay 2013; Jiang et al.\ 2013). The primary mechanism for
the poloidal field generation in this model is the \bl\ mechanism involving the decay of
tilted bipolar sunspots (Babcock 1961; Leighton 1969). Since this mechanism depends on
the existence of sunspots in order to be operative, this mechanism may not work during
a grand minimum when there are no sunspots.  We would then require some other mechanism
to pull the Sun out of the grand minimum. Early models of the solar dynamo invoked
the $\alpha$-effect proposed by 
Parker (1955) and Steenbeck et al.\ (1966) to generate the poloidal field.
The $\alpha$-effect can twist a toroidal field to produce a poloidal field only if 
the toroidal field is not stronger than the equipartition field.  After simulations of the
buoyant rise of flux tubes suggested a much stronger toroidal field (Choudhuri \& Gilman 1987;
Choudhuri 1989; D'Silva \& Choudhuri 1993; Fan et al.\ 1993), the flux transport dynamo models
used the \bl\ mechanism as the favoured mechanism rather then the $\alpha$-effect 
for generating the poloidal field. During a grand minimum, the toroidal field presumably
becomes much weaker and probably the $\alpha$-effect can be operative to pull the Sun
out of the grand minimum. Since we have very little knowledge of the nature of
this $\alpha$-effect, CK12 assumed the same \bl\ mechanism to be operative all the
time to simplify the theoretical calculations. One of the things we explore in this
paper is the nature of the $\alpha$-effect needed to pull the dynamo out of the grand
minimum.  We shall see that various observational requirements put some important
constraints on the nature of this $\alpha$-effect.

Let us now come to the question of what can cause irregularities in the solar
cycle and the grand minima. One important question is whether the nonlinearities
in the system can induce chaotic behaviour. The simplest kind of nonlinearity used
extensively in the earlier dynamo models is the $\alpha$-quenching.  If the magnetic
field becomes stronger than usual due to some reason, this quenching makes $\alpha$
smaller and the dynamo weaker, bringing down the magnetic field. If the magnetic field
becomes weaker, then the opposite happens.  A nonlinearity in the form of 
$\alpha$-quenching makes the dynamo more stable instead of producing chaotic
behaviour. A few authors have found intermittency behaviour in highly truncated 
dynamo models with more complicated kinds of nonlinearity which do not seem
justified by solar observations (Weiss et al.\ 1984; Wilmot-Smith et. al.\ 2005).

One other source of irregularity is stochastic noise. Since the mean-field dynamo
theory is obtained by averaging over turbulence, we expect turbulent fluctuations
to provide a random noise. Hoyng (1988) realized this for the first time and later 
several authors  showed that stochastic noise introduced in the mean-field dynamo equation
can produce irregularities in solar cycles including grand minima-like episodes
(Choudhuri 1992; Charbonneau et al. 2004; G\'omez \& Mininni 2006; Brandenburg \& Spiegel 2008; Moss et al. 2008; 
Usoskin et al. 2009a; Passos et al.\ 2012). A scenario for the origin
of stochastic noise in a flux transport dynamo was provided by Choudhuri et al.\ (2007).
The Babcock--Leighton mechanism for poloidal field generation depends on the tilts of
bipolar sunspots. Although the average tilt at a latitude is given by Joy's law, one finds a scatter
around the average (Dasi-Espuig et al. 2010), presumably due to the effect of convective turbulence
on the rising flux tubes (Longcope \& Choudhuri 2002).  So we believe that the Babcock--Leighton
process intrinsically has a random component.  Choudhuri et al.\ (2007) incorporated this
effect allowing the poloidal field generated at the end of a cycle to differ from its
average value.  This approach has been followed in subsequent papers (Jiang et al.\ 2007;
Goel \& Choudhuri 2009; Choudhuri \& Karak 2009) as well as the present paper.
Choudhuri \& Karak (2009) have shown that, if the poloidal field becomes sufficiently 
weak at the end of a cycle due to fluctuations in the Babcock--Leighton mechanism, then 
that may trigger a grand minimum. 
 
Another source of irregularities is fluctuations in meridional circulation, of which the importance
has been recognized only recently (Yeates et al. 2008; Karak 2010; Karak \& Choudhuri 2011; Passos 2012).
The meridional circulation plays a crucial role in the \ftdm\
(Choudhuri et al.\ 1994; Dikpati \& Charbonneau 1999; Karak 2010).
Although we have neither good theoretical understanding nor long observational measurements of the 
meridional circulation, Karak \& Choudhuri (2011) used durations of sunspot cycles in the last 250 years to have some
idea about fluctuations in \mc.  They concluded that the \mc\ had large temporal variations with
coherence time more than a solar cycle. There are also many other evidences for variations of the \mc\ 
in past (e.g., Passos \& Lopes 2008; Passos 2012).
When the meridional circulation slows down, the period of the dynamo becomes longer.
This has rather different effects on dynamo models with high and low turbulent diffusivity (Yeates et al.\ 2008; Karak 2010; Karak \& Nandy 2012).
If the turbulent diffusivity is assumed reasonably high (which is the case in our model),
then the cycles become weaker because diffusivity has a longer time to act in a cycle.
On the other hand, if the turbulent diffusivity is low (Dikpati \& Charbonneau 1999),
then the effect of diffusivity is not so strong and the cycles become stronger with decreasing meridional circulation
because the differential rotation has a longer time to act on the magnetic fields. Only a dynamo
model with reasonably high turbulent diffusion (like what we use) can explain observational
effects like the dipolar parity of the Sun (Chatterjee et al.\ 2004; Hotta \& Yokoyama 2010), 
the Waldmeier effect (Karak \& Choudhuri 2011), the period and the amplitude relation (Karak 2010) and the lack of significant hemispheric
asymmetry (Chatterjee \& Choudhuri 2006; Goel \& Choudhuri 2009).  See \S~5 of Jiang et al.\ (2007) 
and Miesch et al. (2012) (also see Mu\~noz-Jaramillo et al.\ 2013) 
for a discussion on this topic. In the dynamo model with high diffusivity in which a weaker
\mc\ makes cycles weaker, Karak (2010) has shown that a sufficiently weak \mc\ can trigger a 
grand minimum.

Our recent paper CK12 studies the occurrence of grand minima in our theoretical dynamo model
by introducing simultaneous fluctuations in the poloidal field generation and the \mc.
The levels of fluctuations were determined from the observational data of the last 28 cycles.
With such fluctuations, the \ftdm\ developed in our group showed 24--30 grand minima in a
typical run of 11,000 years---in close agreement with observational data. 

Because of the
shortness of this Letter paper CK12, a full exploration of the different aspects of the problem
could not be presented in it.  This is done in the present paper.   After giving 
a short introduction to the model in \S~2, we
present in \S~3 various aspects of the results not discussed in CK12, such as the relative
importance of the two fluctuations and the dependence on parameters like the coherence time
of the meridional circulation.  We also present some results of grand maxima, which could not be
discussed in CK12.  Finally, \S~4 addresses the important question of how the dynamo comes out
of the grand minimum and if we can say something about the nature of the $\alpha$-effect which
may be needed for this.

\section{Model}
We carry out all the calculations using a \ftdm\ originally presented in Chatterjee et al.\ (2004). 
This model is based on the kinematic mean-field dynamo theory in which the poloidal field generation 
is assumed to be due to the \bl\ process. Assuming axisymmetry, the evolutions of the magnetic field
components in this model are 
described by the following two equations:
\begin{equation}
\frac{\partial A}{\partial t} + \frac{1}{s}({\bf v}.\nabla)(s A)
= \eta_{p} \left( \nabla^2 - \frac{1}{s^2} \right) A + \alpha B,
\end{equation}
\begin{equation}
\frac{\partial B}{\partial t}
+ \frac{1}{r} \left[ \frac{\partial}{\partial r}
(r v_r B) + \frac{\partial}{\partial \theta}(v_{\theta} B) \right]
= \eta_{t} \left( \nabla^2 - \frac{1}{s^2} \right) B 
+ s({\bf B}_p.{\bf \nabla})\Omega + \frac{1}{r}\frac{d\eta_t}{dr}\frac{\partial{(rB)}}{\partial{r}},
\end{equation}
with $s = r \sin \theta$. Here $A$ is the vector potential of the 
poloidal magnetic field ($\bf {B}_p$), $B$ is the toroidal magnetic field, $\vf=v_r \hat{r} + 
v_{\theta} \hat{\theta}$ is the meridional
circulation, $\Omega$ is the internal angular velocity, $\eta_p$, $\eta_t$ are the turbulent diffusivities for the poloidal and toroidal
components, and $\alpha$ is the source term for the poloidal field which parameterized
the \bl\ mechanism. 
%Here $\alpha = \alpha_{\mathrm{BL}}$ and its radial dependence is shown in Figure~\ref{alpha_prof}. 
%In section~\ref{section:recovery} we shall discuss more about the source term $\alpha$. 
The details of all these parameters are specified in Chatterjee et al.\ (2004). However, Karak (2010) recently
modified a few parameters slightly and in this work we are using exactly the same parameters as used in Karak (2010).

Just to remind the readers, we mention that in the expression of the \mc\ there is a parameters
$v_{0}$ which determines the strength of the meridional circulation. For normal cycle with period of
11~years we take $v_0 = 23$~m~s$^{-1}$. However, in this work, when we introduce fluctuations in the meridional circulation, we 
change this $v_0$ to change the strength of the meridional circulation.

Let us make a comment on the absolute value of the magnetic field in our results.  If the equations
are completely linear in the magnetic field, then the unit of the magnetic field would be 
arbitrary.  While (1) and (2) are linear equations, our problem becomes nonlinear when
we include magnetic buoyancy following the methodology used in the earlier papers from our 
group (see Chatterjee et al.\ 2004).
If the amplitude of the toroidal magnetic field $|B|$ above the bottom of the convection zone
is larger than a critical value $B_c$, then a part of the toroidal field is made to rise to surface.
This nonlinearity limits the amplitude of the magnetic field.  We take $B_c = 0.8$ which makes
the maximum value of the magnetic field at the bottom of the convection zone hover around 1.
Since simulations of flux tube rise based on the thin flux tube equation 
(Spruit 1981; Choudhuri 1990) suggest magnetic fields of the order of $10^5$ G at the
bottom of the convection zone (Choudhuri \& Gilman 1987;
Choudhuri 1989; D'Silva \& Choudhuri 1993; Fan et al.\ 1993), it is tempting to identify
the value 1 of the magnetic field in our simulations with $10^5$ G.  However, such an identification
is questionable.  Apart from the fact this would give values of the polar magnetic field disagreeing with
observations, we expect the magnetic field to be $10^5$ G only inside flux tubes, whereas the dynamo
equation deals with the mean magnetic field.  If the filling factor of flux tubes at the bottom
of the convection zone is considerably less than 1, then the mean magnetic field which has to be
identified with magnetic field computed in our model may be much less than $10^5$ G.  We shall
discuss these considerations further in \S~4 where we discuss whether the dynamo requires
an $\alpha$-effect in addition to the \bl\ mechanism to bring it out of the grand minimum.

In the next section, we present results based on exactly the same model of producing grand minima
which was used in CK12.  We basically discuss a few important aspects of the problem which could
not be included in CK12 due to the lack of space.  Then, in \S~4, we shall allow the possibility that
the poloidal field generation under normal cycle conditions and during the \gm\ may require
different mechanisms.

\section{Simulations of grand minima}

Our earlier paper CK12 explained the basic assumptions of our model of grand minima and
presented some illustrative results.  CK12 being a letter paper, a full discussion of
the results could not be presented in it.
This section presents some additional results based on the grand minimum model of CK12.
Since the details of the model have been given in CK12, we simply mention the salient
features. The grand minima in our model were produced by fluctuations in \mc\ and by fluctuations
in the \bl\ process that would make scaled polar field amplitude $\gamma$ at the end of
a cycle vary from cycle to cycle. Assuming the fact that the
solar cycle period is inversely related to the strength of the \mc,
we estimated the nature of fluctuations in the \mc\ from the durations of the last
28 cycles. On the other hand, assuming that the solar cycle strength is directly 
correlated to the strength of the polar field at the end of the previous cycle, we have obtained 
the fluctuations in $\gamma$ from the strengths of the last 28 solar cycles. 
See Figure~2 of CK12 and the corresponding text for details. Assuming both the fluctuations
to obey Gaussian distributions, we have constructed the distributions of these
fluctuations by using the mean and the standard deviations
of these data. Figure~1 shows the typical histograms of these. 
We then make $v_{0}$ and $\gamma$ vary randomly following these
distributions. We point out that
for different runs we generate different results for different realizations of the fluctuations of $v_{0}$ and $\gamma$.
Another important thing to note is that we change the polar field by the factor $\gamma$ at every solar minimum whereas we change $v_{0}$ after a certain time interval called the coherence time $\tau_{MC}$.
Now we shall explore the origin and different aspects of grand minima systematically.

\begin{figure}[!h]
\centering
\includegraphics[width=1.0\textwidth]{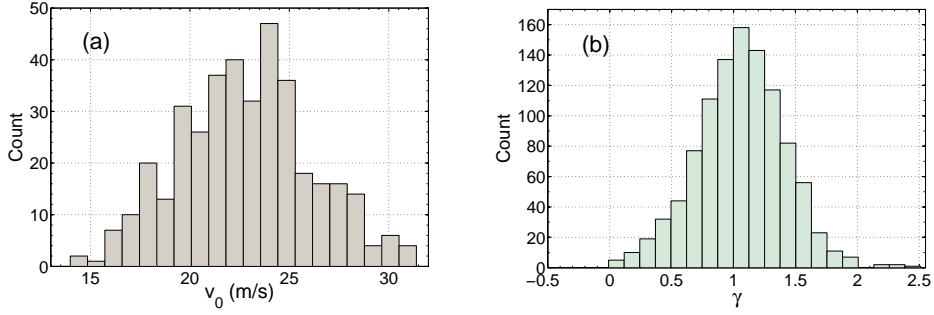}
\caption{Typical histograms of (a) the strength of the \mc\ $v_{0}$ (left plot) and 
(b) the strength of the poloidal field $\gamma$ (right plot) 
used for grand minima simulations. These randomly generated data are taken from Gaussian distributions whose means 
and standard deviations are derived in CK12.}
\label{hists}
\end{figure}

\begin{figure}[!h]
\centering
\includegraphics[width=1.1\textwidth]{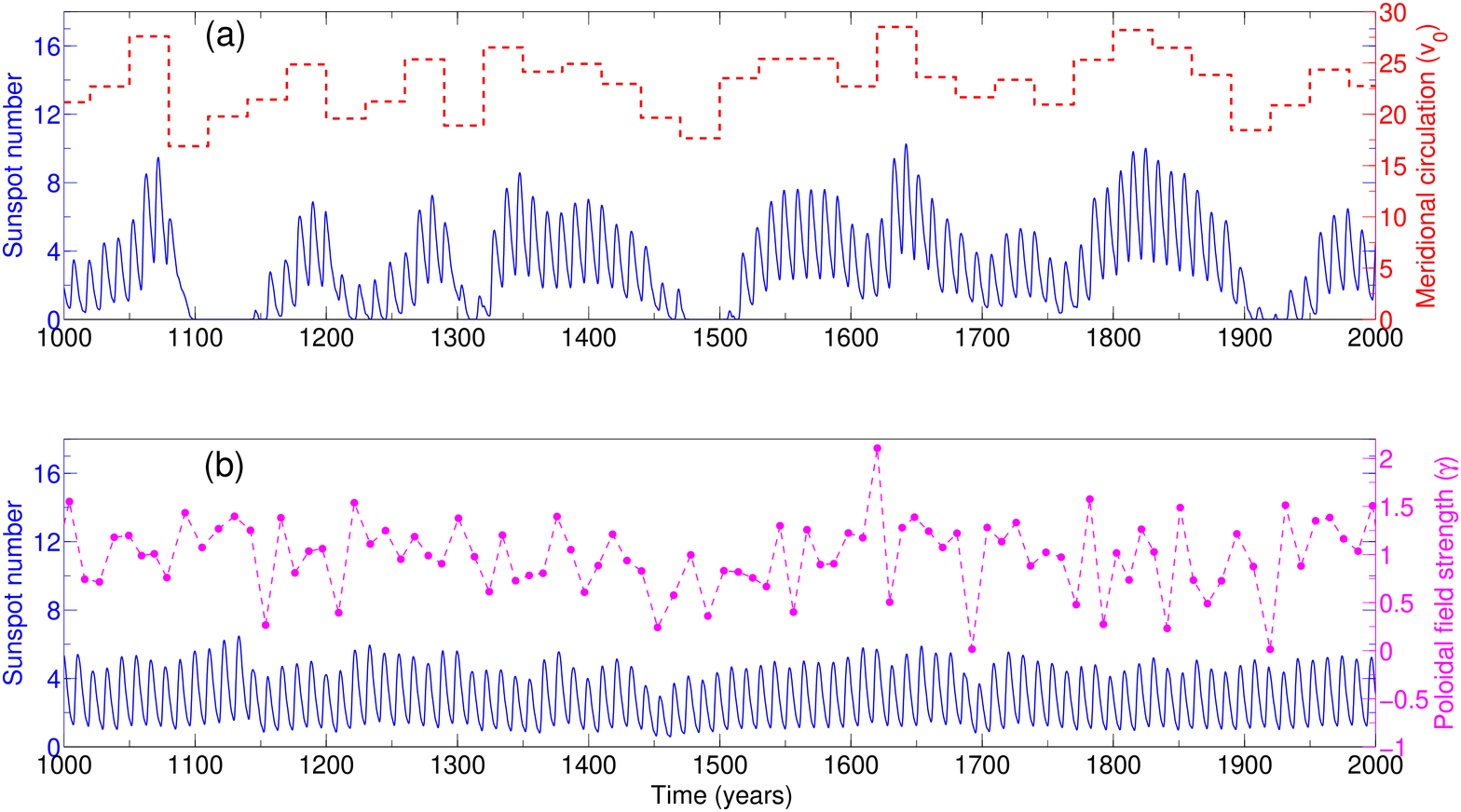}
\caption{(a) Upper panel: Simulation of grand minima with fluctuating \mc\ with $\tau_{MC} = 30$~years. The dashed (red) line shows the strength
of \mc\ $v_{0}$ used in the simulation whereas solid (blue) line shows the (theoretical) sunspot number as a function of time (years). 
Two grand minima are clearly evinced. (b) Bottom panel: Simulation with fluctuating \pf. The dashed (pink) line shows the strength of \pf\
$\gamma$ changed at every solar minimum, whereas solid (blue) line shows the sunspot number.}
\label{ssn_plot}
\end{figure}

\subsection{Contributions of \mc\ and the \pf\ in triggering grand minima}

Using the same parameters of the basic dynamo model as used in CK12, we
explore the relative importance of the contributions of the \mc\ fluctuations and the \pf\ 
fluctuations in triggering grand minima. We perform two separate simulation runs
by including only one kind of fluctuations in each run. 
First, we do a simulation by randomly
varying only $v_{0}$ after intervals of 30~years (i.e., $\tau_{MC}$ = 30~years). The 
random values for $v_{0}$ used for this run have been shown in Figure~\ref{hists}(a).
The top panel of Figure~\ref{ssn_plot} shows the results of this run. For the clarity of display we show only a small clip of 1000~years
long data. Next, we present a simulation with only poloidal field fluctuations. We change the poloidal field factor $\gamma$ at every solar minimum.
The histogram of $\gamma$ for this run is shown in Figure~\ref{hists}(b), whereas the results of this run is shown in Figure~\ref{ssn_plot}(b).
%Finally we run a simulation with these combined fluctuations.

Based on these simulations, we make the following important conclusions.\\
{\it i}) The \mc\ plays an important role in modulating the solar cycle period and the amplitude (consistent with Karak 2010). This is clear from Figure~\ref{ssn_plot}(a).\\
{\it ii}) Most of the \gm\ are produced when the \mc\ becomes sufficiently weak, whereas the weak \pf\ has very minor contribution in
producing \gm\ (compare two panels in Figure~\ref{ssn_plot}). 
%We see that in our final simulation
%with combined fluctuations of \mc\ and \pf, all \gm\ are mainly produced due to the \fl\ in \mc\ alone. 
Fluctuations in 
\pf\ has some effect in producing \gm\ when combined with fluctuating \mc, but are not able to produce any grand minima separately. 

Although we ourselves had not been aware of these conclusions at the time of writing our earlier paper CK12,
it now seems from hindsight that these conclusions would follow from
Figure~1 of CK12, where the region of the parameter space required for producing 
\gm\ is demarcated. When the polar field is not varied (i.e.\ $\gamma$ is always kept equal to 1),
we see from Figure~1 of CK12 that
it is necessary to make $v_0$ less than about 15.5 m s$^{-1}$ to produce \gm\ and we find
that the probability of this is not too low, as can be seen from Figure~2(a) of CK12.  On the other hand, if fluctuations in \mc\ are
not included (i.e.\ $v_0$ is held fixed at the value 23 m s$^{-1}$), then $\gamma$ has to be made about $-1$
(below the bottom of Figure~1 of CK12) and we see from Figure~2(b) of CK12 that the probability of this is
miniscule.

Now we come to the question whether our conclusion that the fluctuations in the polar field
do not play an important role in producing grand minima changes when the parameters of the
basic dynamo model are different.  This is discussed in the next subsection.

\subsection{Sensitivity of the results on the value of $\alpha$}

\begin{figure}[!h]
\centering
\includegraphics[width=1.05\textwidth]{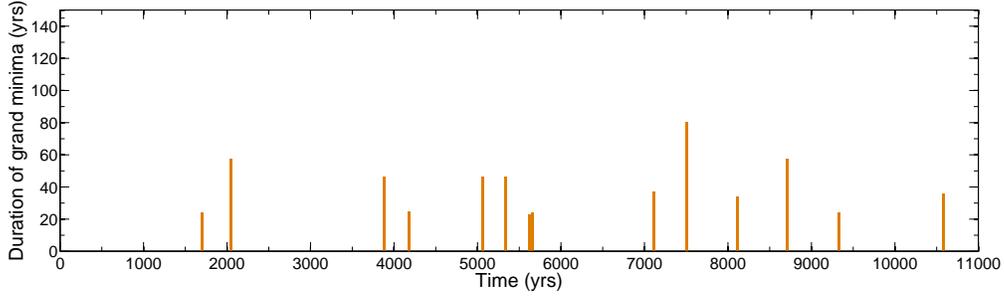}
\caption{The durations of grand minima indicated by vertical bars at their times of occurrence in a 11,000~yr simulation with only \pf\ fluctuations.
Note that unlike earlier in this case we get 14 grand minima with \pf\ fluctuations alone. In this simulation $\alpha_0 = 24$ m~s$^{-1}$ instead of 30 m~s$^{-1}$ used earlier; everything else remain unchanged.}
\label{gm_gamma}
\end{figure}

When we run our basic dynamo model without fluctuations, we find that the critical value of
the $\alpha$ coefficient is $\alpha_0 = 21.1$ m~s$^{-1}$.  In other words, when we run the
code by varying $\alpha_0$ alone and keeping all the other parameters fixed (especially using
the diffusivity $\eta_0 = 3 \times 10^{12}$ cm$^2$~s$^{-1}$), we get non-decaying oscillatory
solution only if $\alpha_0$ is larger than this critical value.  The results of CK12 as well
as the results presented in \S~3.1 are based on a model using a somewhat supercritical
value $\alpha_0 = 30$ m~s$^{-1}$. We now carry on some calculations using only a moderately
critical value $\alpha_0 = 24$ m~s$^{-1}$. We study the effect of introducing fluctuations
in polar field alone on this moderately critical dynamo. Figure~\ref{gm_gamma}
shows a typical result of such a simulation. Now we find that 14 grand minima are produced in 11,000 yr.
It is not difficult to give a physical argument why fluctuations in polar field produce
grand minima more easily in a moderately critical dynamo.  If the fluctuations make the polar field during a
minimum much less than the polar field which such a moderately critical
dynamo would normally produce, then this is almost as if the strength of $\alpha$ is temporarily
reduced and this can make the dynamo subcritical, pushing it into a grand minimum. This is
not likely to happen when the dynamo is reasonably supercritical. We thus conclude that
fluctuations in the polar field would have a significant effect on the dynamo only
if it is moderately critical.
 
We have done some calculations introducing 
fluctuations in both the \mc\ and the \pf\ in a moderately critical dynamo.
The number of \gm\ becomes much larger than the observed value. 
Since the results of a supercritical dynamo are in such good agreement
with the observations (CK12), one is tempted to conclude that the solar dynamo
is supercritical and fluctuations in the polar field do not have much
effect in inducing grand minima. It may be noted that Charbonneau et al.\ (2007)
proposed that the Gnevyshev--Ohl effect is produced by period doubling,
which also requires a reasonably supercritical dynamo. Other dynamo models
which are not supercritical show the occurrence of grand minima on
introducing fluctuations in $\alpha$-effect alone (Olemskoy et al.\ 2013).

We point out that the earlier paper by Choudhuri \& Karak (2009) produced grand minima only
with fluctuations in the \bl\ mechanism.  The model used in this paper was not too
supercritical.  Additionally, one ad hoc assumption used in this paper was to reduce
the toroidal field also by a factor 0.8 when the poloidal field was reduced to create a grand minimum.
This helped in creating the Maunder-like grand minima.  In the present paper, we do not use this ad hoc assumption
and the toroidal field is never changed when changing the poloidal field to incorporate fluctuations
in the \bl\ process.

The results presented in the remaining subsections of this section are all
obtained with the supercritical dynamo used in CK12.

\subsection{Are initiations of \gm\ sudden or gradual?}
 
One important question connected with grand minima is whether they initiate suddenly or gradually.
Usoskin et al.\ (2000) concluded that the Maunder minimum started abruptly.  However, Vaquero et al.\
(2011) now present evidence that the initiation of the Maunder minimum was more gradual. On the
theoretical side, in our simulation runs we find both \gm\ which start abruptly and which start
gradually.  Since we now recognize the fluctuations in \mc\ to be the primary cause of \gm, we
discuss the results for the run with fluctuations in \mc\ alone.  Things can be seen more
cleanly there.  The results are qualitatively the same when both fluctuations in \mc\ and polar
field are present.

In Figure~\ref{ssn_plot}(a) we see that the grand minimum that started slightly before 1500 and also
the grand minimum that started around 1900 initiated gradually. 
These \gm\ did not start immediately after the \mc\ became sufficiently weak.
The dynamo took about one or two solar cycles to enter into \gm. Therefore, 
one or two solar cycles before the beginning of 
\gm, the solar cycle period tends to become longer (because the \mc\ determines the cycle period). 
This result is remarkably consistent with the results of Miyahara et al. (2010) who have found sufficient evidences of the  
longer solar cycles even before the beginning of the \mm\ and also the Sp\"orer minimum.

Now we discuss an opposite case where the grand minimum starts suddenly and we do not see much change of the solar cycle 
period before the beginning of the grand minimum.  In Figure~\ref{ssn_plot}(a), the grand minimum around 1100 shows this
behaviour.  We note that this grand minimum was caused by the \mc\ falling to a rather very low value
suddenly from a reasonably high value. Although the cycle period did not get elongated before the
start of the grand minimum, the cycle period is longer than the usual during the actual grand minimum epoch.

Based on our theoretical results, we draw the following conclusion.  As the \mc\ is made to fluctuate randomly,
it would sometimes happen that the \mc\ would drop from a rather high value to a low value.  In such a situation,
the initiation of the grand minimum seems abrupt.  More commonly, we may have the \mc\ dropping from a more
moderate value to a low value.  The grand minimum starts more gradually in this situation.
As we do not know at present how rapidly the \mc\ can drop to a low value, we do not know which one is more physical.
In our simulations done with both kinds of fluctuations present, 
we have noted that there are about $40\%$ grand minima which initiate abruptly whereas the remaining grand minima initiate
more gradually.  
However the recovery from
\gm\ is always gradual, which is consistent with observations during the \mm\ (Usoskin et al. 2000).
%--------------------------------------------------------------------------------------
\begin{figure}[!h]
\centering
\includegraphics[width=1.0\textwidth]{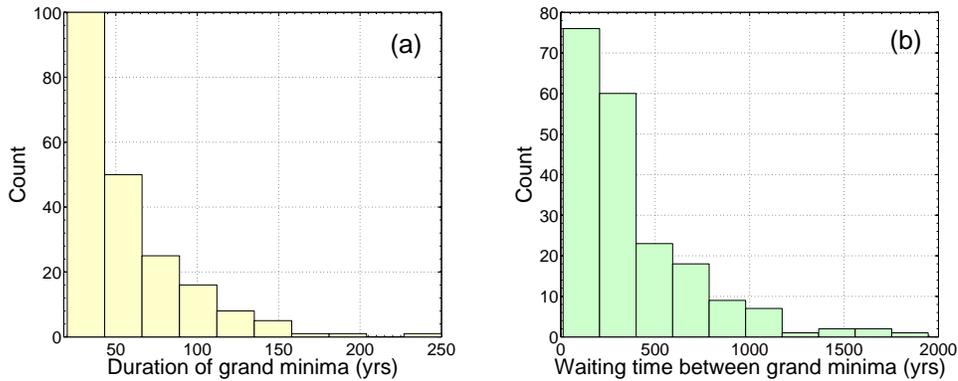}
\caption{Left panel (a) shows the distribution of the durations of the grand minima
and the right panel (b) shows the distribution of the waiting times between the grand
minima. This figure is produced from the data of 88,000~years of simulation run.}
\label{bl_histograms}
\end{figure}
%-----------statistic of grand minima---------------------------------------------------

\subsection{Statistics of grand minima}

In Figure~5 of CK12 we presented the distributions of durations of grand minima and
the waiting times between them.  Since these distributions were constructed from
the limited set of 29 \gm\ which occurred during one run, the nature of these
distributions was not very clear from this figure.
To make a statistically reliable conclusion, we now make histograms from the data of a very long 
simulation (about 88,000~years) in which we have detected about 207 grand minima.
The histograms are shown in Figure~\ref{bl_histograms}.  
From this plot, we now clearly see that both the duration and the waiting time follow exponential behaviour.
This tells that the duration and waiting time are governed by stationary memoryless stochastic processes. As the grand minima 
are produced by the random fluctuations in the \mc\ and the \pf, the occurrence of the grand minimum
must be random events implying that the waiting time distribution is exponential. On the other hand, 
once the dynamo enters into a 
grand minimum state, the recovery of dynamo from \gm\ state is only possible by the increase of \mc\ 
which happens randomly. Since stochastic fluctuations in \mc\ are responsible 
for bringing back the dynamo into the normal cycle, the distribution of the durations of \gm\ is also exponential.
We mention that the observational distribution of the waiting times of \gm\ based on 27 grand 
minima in last 11,400~years reported by Usoskin et al. (2007) is also exponential, whereas the 
distribution of durations is not so conclusive.

%-------------dependence of coherence time -----------------------------------------------
\begin{figure}[!h]
\centering
\includegraphics[width=0.7\textwidth]{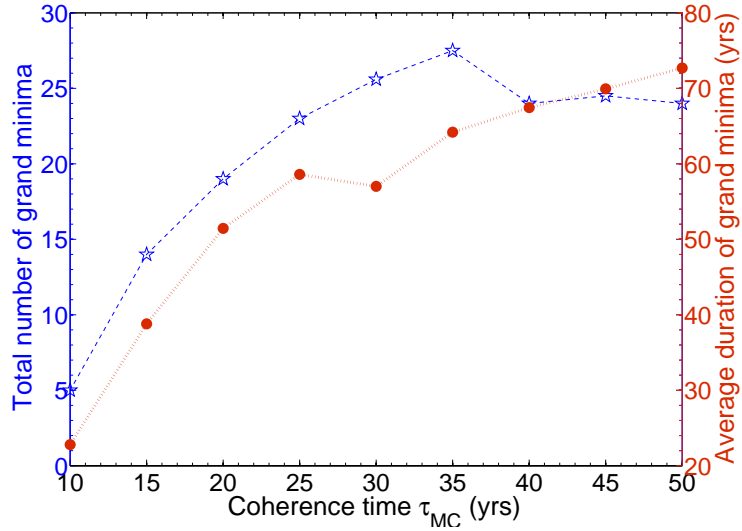}
\caption{Dependence of the number of grand minima (shown by star and dashed line) in 11,000~years and 
their average duration (shown by circle and dotted line) along right vertical axis as a function of the 
coherence time of the meridional circulation ($\tau_{\mathrm{MC}}$).}
\label{tau_depen}
\end{figure}

\subsection{The dependence of the coherence time of meridional circulation}

Karak \& Choudhuri (2011) pointed out that several successive cycles in the past often had very
similar periods (see their Figure~2).  This suggests that the \mc\ probably had remained steady
during those cycles before changing abruptly at the end of such an epoch.  Given the limited data
of the last few cycles, it is very difficult to estimate the coherence time $\tau_{MC}$ of the \mc.
Karak \& Choudhuri (2011) concluded that this coherence time should lie in the range between
15 years and 45 years.  All the results presented in CK12 were obtained by using a coherence
time of 30 years.
Here we explore the importance of this coherence time $\tau_{MC}$ of the \mc\ on various features of grand minima.

In our earlier simulations of CK12 using $\tau_{MC} = 30$~years, after every 30~years $v_0$ 
was varied randomly in accordance with their distributions.
Now we have performed several simulations by varying $\tau_{MC}$ from 10 to 50~years. We note the number of 
\gm\ in  a run of 11,000~years and also calculate the average duration of such \gm\ for each run
with a particular value of $\tau_{MC}$. Note that in all simulations
the \pf\ is changing at every solar minimum only. Figure~\ref{tau_depen} shows the results.
We see that the total number of \gm\ initially increases with the increase of $\tau_{MC}$ and then,
after a certain value around 30~years, it tends to decrease. This is easy to understand.
If $\tau_{MC}$ is small, the dynamo does not get much time to make magnetic fields sufficiently weak
even when the \mc\ falls to a low value during a short coherence time. Therefore,
the occurrence of grand minima becomes less. With the increase of $\tau_{MC}$, the number of \gm\ 
increases. However, after $\tau_{MC}$ becomes comparable to the typical duration of a few cycles,
further increase of it does not increase the number of grand minima rather it decreases. 
If the \mc\ changes after long times, then the probability that it falls to a sufficiently 
low value also becomes less. Therefore the total number of grand minima in a finite period of run 
becomes less at large $\tau_{MC}$.

Another important result from this figure is that the average duration of \gm\ is an increasing function of
$\tau_{MC}$. In an earlier work with constant \mc\ where \gm\ were produced by fluctuations
in polar field alone, Choudhuri \& Karak (2009) found that  
the recovery to the normal state from a grand minimum is only determined by the dynamo 
growth rate (measured by the dynamo number $\sim \alpha/\eta^2$; where $\alpha$ is the strength of the poloidal 
field generation process during grand  minimum episode). However, in the case of fluctuating \mc, where \gm\ are 
mostly caused by the weak \mc, the recovery from grand minima state is not only determined by the dynamo number 
but also on how rapidly the \mc\ comes back to a more usual value from the very low value that caused the
grand minimum. If the \mc\
recovers quickly from its low value, the duration of the grand minimum
will be short and vice versa. Therefore the average duration of 
grand minima is strongly dependent on how frequently the \mc\ changes, i.e., on $\tau_{MC}$.

%%%%%%% -----------------------------  Grand MAXIMA  ----------------------------------
\subsection{Grand maxima}
Of late, grand maxima---epochs during which solar activity becomes exceptionally strong for sufficiently long 
time---are drawing more and more attention of solar physicists. The middle of the twentieth century was  
such an epoch when successive several solar cycles were rather strong (Solanki et al.\ 2004). Along with
grand minima, Usoskin et al.\ (2007) also presented a study of grand maxima during the
last 11,400 years.  They identified 19 grand maxima and showed that their durations follow an exponential
distribution, suggesting that the durations are determined by a memoryless random process.

\begin{figure}
\centering
\includegraphics[width=1.0\textwidth]{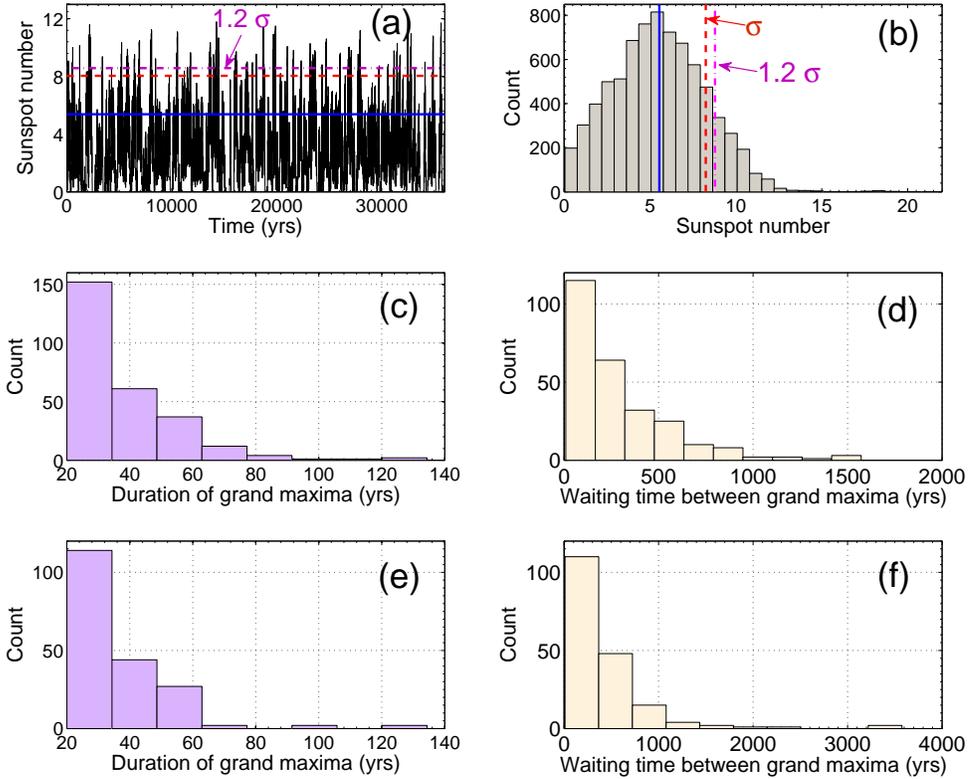}
\caption{Statistics of grand maxima. (a) Shows the theoretical sunspot number. The three horizontal lines (solid, dashed and dash-dotted) 
indicate the values $SN_m$, $SN_m + \sigma$ and $SN_m + 1.2\sigma$, 
where $SN_m$ and $\sigma$ are the mean and the standard deviation of the peak sunspot numbers.
(b) Shows the histogram of the peak sunspot numbers where the three vertical lines (solid, dashed and dash-dotted) 
indicate the values $SN_m$, $SN_m + \sigma$ and $SN_m + 1.2\sigma$.
(c) and (d) Show the histograms of durations and waiting times of grand maxima defined to be above the
$SN_m + \sigma$ level. (e) and (f) Show histograms similar to (c) and (d), except that the grand
maxima are now defined to be above the
$SN_m + 1.2\sigma$ level.}
\label{grand_maxima}
\end{figure}

We present a study of grand minima from our theoretical simulation. How you define grand maxima has more
arbitrariness compared to how you define grand minima. We now explain how we select our grand maxima.
Figure~\ref{grand_maxima}(a) plots the theoretical sunspot eruptions in the numerical run, whereas Figure~\ref{grand_maxima}(b) is a
histogram showing the distribution of the peak values of these cycles. The solid horizontal line in Figure~\ref{grand_maxima}(a)
and the solid vertical line in Figure~\ref{grand_maxima}(b) indicate the mean value $SN_m$ of the cycle peaks. We calculate the
standard deviation ($\sigma$) of these peak values. The dashed (red) lines in Figure~\ref{grand_maxima}(a) and Figure~\ref{grand_maxima}(b) 
indicate the  $SN_m + \sigma$ levels.  If at least two successive solar cycles have their strengths
above this level, then we take it as a grand maximum. Figure~\ref{grand_maxima}(c) and Figure~\ref{grand_maxima}(d) respectively are
the histograms showing the distribution of the durations and the waiting times of these
grand maxima.  To figure out the extent to which the statistical distributions depend on the 
definition of grand maxima, we also present results by defining grand maxima as at least two successive
cycles having peaks above $SN_m + 1.2 \sigma$ levels.  These levels are indicated in Figure~\ref{grand_maxima}(a) and 
Figure~\ref{grand_maxima}(b) by the dash-dotted (pink) lines.  The distributions of durations and waiting times for grand maxima
defined in this way are shown in Figure~\ref{grand_maxima}(e) and Figure~\ref{grand_maxima}(f). Comparing them with Figure~\ref{grand_maxima}(c) and Figure~\ref{grand_maxima}(d),
we conclude that the statistical behaviours of grand maxima are reasonably robust and do not change with the definition
of grand maxima. 
 
We remind the reader that, to get a statistically significant result, we used data of about 88,000~years long run.
The numbers of grand maxima were 270 and 191 in the two definitions. This implies that the numbers of grand maxima
in 11,000 years would be 34 and 24.   
Interestingly the distributions of durations and the waiting times of grand maxima 
as shown in Figure~\ref{grand_maxima}(c)--(d) also seem to follow the exponential distribution, which again tells us that the
occurrence of grand maxima is governed by a stationary memoryless random process.  On comparing with
Figures~\ref{bl_histograms}(a)--(b) showing similar distributions for grand minima, we note that the durations of grand
maxima tend to be shorter than the durations of grand minima.  A physical explanation for this is not
difficult to give.  Once the Sun enters a grand minimum due to an effect like the slowing down of the
\mc, the dynamo has to build up the magnetic field
again before the Sun can come out of the grand minimum even after the \mc\ has returned to more normal
values.  This takes some time.  On the other hand, from a state of grand maximum, the activity level
can get reduced more easily when conditions change due to stochastic fluctuations.
Another important point to note is that like \gm\ grand maxima are mainly caused by the
variable meridional circulation. This becomes clear by looking at Figure~\ref{ssn_plot}(a). We see
that strong meridional circulation makes the cycle stronger.

%%%%%%%%%%%%%%%%%%%%%%%%%%%%%%%%%%%%%%%%%%%%%%%%%%%%%%%%%%%%%%%%%%%%%%%%%%%%%%%%%%%%%%%%%%%%%%%%%%%
% ------------------------------ Alpha effect during grand minima----------------------------------
\section{The recovery phase of grand minima}
     \label{section:recovery}

\begin{figure}[!h]
\centering
\includegraphics[width=1.0\textwidth]{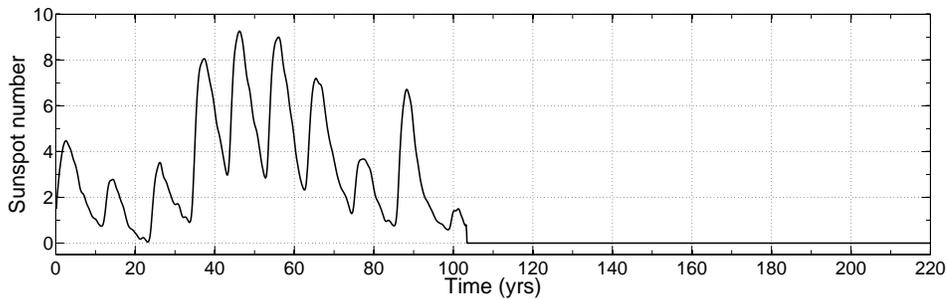}
\caption{Simulation of the solar cycle with the Babcock-Leighton mechanism, which gets switched
off when there are no sunspots.  The dynamo is not able to recover from the grand minimum state.}
\label{bl_cutoff}
\end{figure}

Here we explore an important, yet unsettled, issue connected with \gm: the recovery mechanism from grand minima states.
If the \mc\ or the \pf\ somehow becomes sufficiently weak, then that can push the Sun
into a grand minimum. However, we do not understand well how the Sun comes out of such
a quiescent state. There are also many uncertainties in our understanding of 
the nature of the dynamo process during the grand minimum state.
It has been clearly demonstrated by observations 
that the solar cycle continued during grand minima with  
weaker strength and also with polarity reversals (Fligge et al. 1999; Miyahara et al. 2004, 2010; Nagaya et al. 2012).
The question that remains open is how the \pf\ is generated during grand minima. The \bl\ process depends 
on the decay of tilted active regions and at present 
we have strong observational evidence that this process is indeed working near 
the solar surface (Dasi-Espuig et al.\ 2010; Kitchatinov \& Olemskoy 2011; Jiang et al.\ 2013).
On the other hand, very few sunspots were detected during the \mm\ (Sokoloff \& Nesme-Ribes 1994;
Hoyt \& Schatten 1996). Therefore, the \bl\ process may have been ineffective
during the Maunder minimum. However, in all our earlier 
calculations (Choudhuri \& Karak 2009; Karak 2010; CK12), we have used the same $\alpha$ 
concentrated near the solar surface corresponding to the \bl\ mechanism all the
time for the \pf\ generation because of our lack of knowledge about handling the problem in a better
way. Now we assume that the \bl\ process cannot operate when the toroidal field is very weak
and sunspots do not form. We make the \bl\ $\alpha$ coefficient fall to zero when the dynamo
enters a grand minimum and keep running the simulation.  Then the dynamo cannot come out of
the grand minimum state, as seen in Figure~\ref{bl_cutoff}. 

If we assume that the \bl\ process cannot work during a grand minimum, we need some other mechanism
to pull the dynamo out of the grand minimum. The obvious other candidate to produce the \pf\ is the $\alpha$-effect 
based on helical turbulence ($\alpha_{\mathrm{HT}}$) proposed by
Parker (1955) and Steenbeck et al.\ (1966). This is a mechanism of generating the \pf\ in the convection zone by the twist of the helical turbulence 
which is effective in weak toroidal field regime. When flux tube simulations showed that sunspots form from
toroidal magnetic fields as strong as $10^5$ G (Choudhuri \& Gilman 1987; Choudhuri 1989; D'Silva \& Choudhuri 1993;
Fan et al.\ 1993), the \bl\ mechanism was favoured over the $\alpha$-effect in the flux transport dynamo model, since
the helical turbulence cannot twist toroidal fields as strong as $10^5$ G. Presumably the situation gets reversed
during a grand minimum.  While the \bl\ mechanism may not be operative due to the lack of sunspots, the weaker
toroidal magnetic field during the grand minimum may allow  the $\alpha$-effect to work. We now present some simulations
in which we switch on an $\alpha$-effect during the grand minima and investigate what conclusion we can draw about
the nature of the $\alpha$-effect by requiring that results of the simulation agree with observations. 
These simulations pertaining to the plausible failure of the \bl\ dynamo to recover from a grand minimum 
episode and the role of an additional mean-field alpha effect in a possible recovery are motivated by
discussions with Dibyendu Nandy and Soumitra Hazra (private communications) and the work detailed in 
Passos (2010, PhD Thesis), Hazra, Passos and Nandy (2013) and Passos, Hazra and Nandy (2013).
One other 
point needs to be noted.  Because of the way we treat magnetic buoyancy in our code, whenever the toroidal field
strength above the bottom of the convection zone exceeds $B_c$, a part of it is brought to the solar surface.
During the usual situation (i.e.\ outside grand minima), the toroidal field near the surface in our simulation
continuously gets enhanced by magnetic buoyancy.  This does not happen during the grand minima.  Whether we allow
the \bl\ process to continue or replace it by the $\alpha$-effect, the toroidal field at
the solar surface during grand minima comes there due to turbulent diffusion or advection due to the \mc\ after
being created in the tachocline.

\begin{figure}[!h]
\centering
\includegraphics[width=0.5\textwidth]{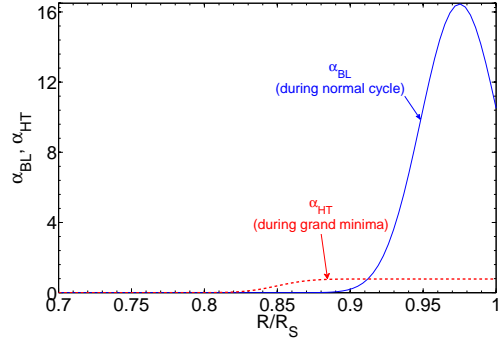}
\caption{Variations of the strength of the Babcock-Leighton $\alpha_{\mathrm{BL}}$ (solid line) and the (helical) 
turbulent $\alpha_{\mathrm{HT}}$ (dashed line) 
as a function of solar radius at $45^0$ latitude.}
\label{alpha_prof}
\end{figure}

As soon as the dynamo enters into a grand minimum state and the sunspot eruption stops 
due to the weak toroidal field, we switch off the $\alpha_{\mathrm{BL}}$ corresponding to
the \bl\ mechanism and switch on the $\alpha_{\mathrm{HT}}$ representing the twisting of the toroidal field
by helical turbulence.
Then, after the recovery from the grand minimum state, we switch off 
the $\alpha_{\mathrm{HT}}$ and again switch on the $\alpha_{\mathrm{BL}}$.
Although the nature and also the sign of this $\alpha_{\mathrm{HT}}$ is not certain at present, we use the 
following profile for it:
 $$\alpha_{\mathrm{HT}} = 1.1~ \mathrm{cos} \theta ~ \frac{1}{2}[1 + \mathrm {erf}( \frac {r - 0.85\Rs} {0.025\Rs})]~ \mathrm{m~s}^{-1}\eqno(3)$$
The profile of this $\alpha_{\mathrm{HT}}$ along with the \bl\ $\alpha_{\mathrm{BL}}$ is shown in Figure~\ref{alpha_prof}. Note that
$\alpha_{\mathrm{HT}}$ is almost one order of magnitude smaller than the $\alpha_{\mathrm{BL}}$ and 
importantly $\alpha_{\mathrm{HT}}$ is zero below around $0.8\Rs$. We have seen that, 
if $\alpha_{\mathrm{HT}}$ is non-zero within the whole body of the 
convection zone, then the solar cycle periods during \gm\ become very short, which is not supported by the
observation (Fligge et al. 1999; Miyahara et al. 2004, 2010; Nagaya et al. 2012). One important conclusion
we draw is that an $\alpha$-effect which gets switched on during grand minima has to be restricted in
the upper regions of the convection zone if we do not want the periods to become too short.

\begin{figure}[!h]
\centering
\includegraphics[width=1.0\textwidth]{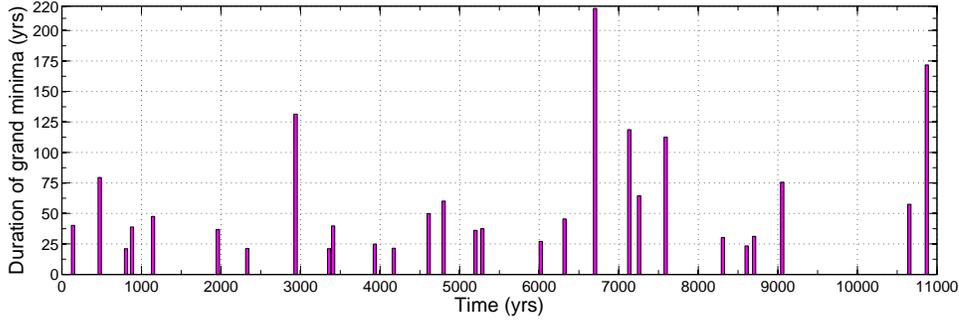}
\caption{The durations of grand minima indicated by vertical bars at their times of occurrence in a 11,000~yr simulation.
This is the result of a particular realization of random fluctuations that produced 28 grand minima. In this simulation,
during \gm\ episodes, the \bl\ $\alpha_{\mathrm{BL}}$ is switched off and a weak turbulent $\alpha_{\mathrm{HT}}$ is allowed to switch on with the
amplitude 1.1~cm~s$^{-1}$.}
\label{alpha_gm}
\end{figure}

\begin{figure}[!h]
\centering
\includegraphics[width=1.0\textwidth]{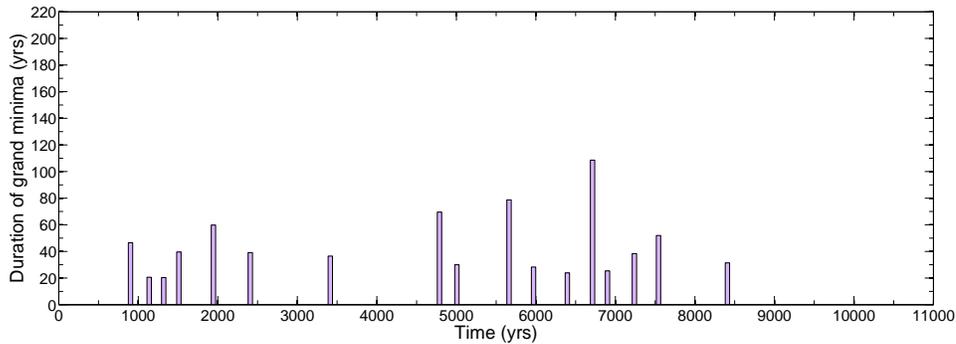}
\caption{The same as Figure~9 except that turbulent $\alpha$ switched on during the grand minima has the
amplitude 1.2~cm~s$^{-1}$.}
%\label{alpha_gm}
\end{figure}

\begin{figure}[!h]
\centering
\includegraphics[width=1.0\textwidth]{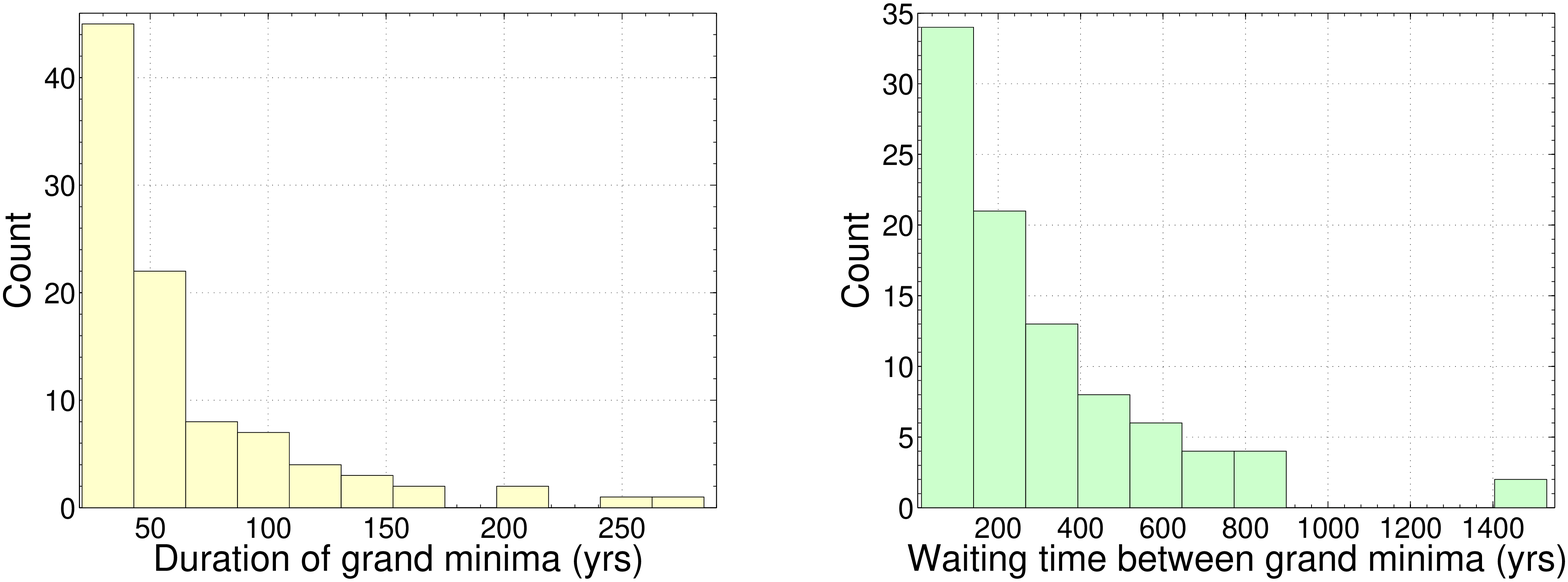}
\caption{Same as Figure~\ref{bl_histograms}, except that the turbulent $\alpha_{\mathrm{HT}}$ gets switched on during
grand minima instead of \bl\ $\alpha_{\mathrm{BL}}$ operating all the time.}
\label{alpha_hist}
\end{figure}

Now let us comment on the amplitude of $\alpha_{\mathrm{HT}}$.  If $\alpha_{\mathrm{HT}}$ has the amplitude 1.1~m~s$^{-1}$
in the upper part of the convection zone which follows from (3), 
the results of the simulation are qualitatively exactly similar to the results
we got by allowing the \bl\  $\alpha_{\mathrm{BL}}$ to operate all the time.  
The results of the simulations of the \gm\ with this $\alpha_{\mathrm{HT}}$ during grand minima is shown in Figure~\ref{alpha_gm}.
This plot shows the positions of the grand minima along the time axis whereas the vertical axis shows the durations of grand minima.
In this 11,000~years simulation run, we get about 28 grand minima. A comparison with Figure~3 of CK12 shows that
the results are qualitatively very similar.
Figure~10 shows the results when $\alpha_{\mathrm{HT}}$ is made to have 
the slightly larger amplitude 1.2 m~s$^{-1}$.
We see that the number of grand minima is reduced in this situations.   When $\alpha_{\mathrm{HT}}$ is
larger, the dynamo gets out of the state of reduced activity very quickly.  Since we count something as grand minimum
only if two successive cycles are missed, the number of grand minima is reduced.  We also did runs
by using the slightly lower value of the amplitude 1.0 m~s$^{-1}$ for $\alpha_{\mathrm{HT}}$.
In this case, we found that the dynamo was unable to get out of a grand minimum after entering it.
Figure~\ref{alpha_hist} presents 
histograms of the durations and the waiting times of grand minima from 33,000~years long simulation data using
the value of $\alpha_{\mathrm{HT}}$ given by (3), with amplitude 1.1 m~s$^{-1}$.
Again in this case both the distributions are exponential and qualitatively similar to the distributions
shown in Figure~\ref{bl_histograms}.

We are not completely sure what conclusions we should make out of the results we have presented. It is quite
remarkable that the results of our simulation by using the same \bl\ $\alpha$ all the time, as had been done
in CK12 and in \S~3 of this paper, are in such good agreement with different aspects of observational data.
On the other hand, if we switch off this \bl\ $\alpha$ concentrated near the solar 
surface during the grand minima and use the traditional
$\alpha$-effect to pull the dynamo out of the grand minima, then we have to fine-tune the nature of this
$\alpha_{\mathrm{HT}}$ quite a bit in order to get results consistent with observational data.  
Interestingly, we get the best results when the amplitude of the $\alpha$-effect is just marginally
above the critical value and only operating in the upper half of the convection zone.  Does this
tell us that \bl\ $\alpha$ remains operative even during the grand minima for reasons we do not understand
now? This is a very provocative question which needs further investigation.
Smaller active regions with
magnetic flux less than that of detectable sunspots may have some (statistical) systematic tilt to produce a significant
\pf\ during \gm-like episodes (see the discussion in Wang \& Sheeley 2013). In fact, Stenflo \& Kosovichev (2012) find a systematic tilt for a long range 
of the magnetic fluxes in active regions, suggesting that the \pf\ may be generated when there are no detectable sunspots. Importantly, even a few big sunspots (with correct tilt) can produce a significant poloidal field to maintain the polarity reversal---this might
also be the case during \gm. As we do not have sufficient observational study of all these issues during the \mm, 
we cannot conclusively say anything about the \pf\ generation mechanism during \gm.
However Passos (2010, PhD Thesis), Passos et al.\ (2013) and Hazra et al.\ (2013) believe the \bl\ process cannot operate during the grand minimum episode 
and a weak mean-field $\alpha$ in the whole convection zone is needed to recover the Sun out of such phase based on their 
simulations with \ftdm\ and low order time delay dynamo model.

\section{Conclusion}
The aim of the present paper is to follow up our earlier paper CK12 in exploring whether different
aspects of grand minima can be explained on the basis of the flux transport dynamo model of the solar
cycle.  Following earlier work done by our group, we keep using a reasonably high turbulent diffusivity,
which implies that a slowing down of the \mc\ results in longer and weaker cycles.  In such a model,
the \mc\ plays a profound role in producing irregularities of the solar cycle and also the grand
minima. One of the main uncertainties in theoretical models at the present time is our lack of
understanding of the \mc, either from the theoretical or the observational viewpoint. While we do not
yet have a complete theory of the solar \mc, we believe that the turbulent stresses in the solar convection
zone drives it and hence we assume the \mc\ to be confined within the solar convection zone. The 
poleward \mc\ near the solar surface causes the advection of the poloidal field to higher latitudes
(Wang et al.\ 1989; Dikpati \& Choudhuri 1994, 1995). We need an equatorward counterflow at the bottom
of the convection zone for flux transport dynamo to produce proper butterfly diagrams (Nandy \&
Choudhuri 2002). So far we do not have direct observational measurements of this counterflow.  While
helioseismology has been able to provide information about \mc\ in the upper layers of the convection
zone (Giles et al.\ 1997; Braun \& Fan 1998), extracting unambiguous information about \mc\ in the
deeper layers has remained a challenge (Gough \& Hindman 2010; Zhao et al.\ 2012). Any possible
periodic modulations of the \mc\ with the solar cycle is not expected
to produced sustained irregularities of the cycle (Karak \& Choudhuri 2012; Passos \& Lopes 2012). But random fluctuations
of the \mc\ with coherence times longer than solar cycle periods, 
as suggested by the data of past cycles, can have profound
effects on the dynamo.  

Our earlier paper CK12 suggested that grand minima are produced by combined fluctuations in the 
\mc\ and in the \bl\ mechanism for generating the poloidal field.  While our further calculations
support this broad scenario, we now find that the fluctuations in the \mc\ are more important in producing
the grand minima (cf. \S~3.1). 
From the theoretical viewpoint, such variations of \mc are not surprising. We know that the \mc\ is mainly generated from the 
imbalance between two large terms -- the non-conservative part of the centrifugal force and the baroclinic torque 
(i.e., the deviation from the thermal wind balance) (Kitchatinov and R\"udiger 1995). 
It is not only the case that there is a deviation 
which produces the \mc\ in the solar convection zone but also this deviation fluctuates because of the 
fact that the differential rotation is produced by turbulent convection and the fluctuations in it is 
unavoidable (Brun et al. 2010). This physics already been explored by a mean-field model of Rempel (2005). 
He introduced random fluctuations in the $\Lambda$-effect and found that it produced fluctuations in the differential 
rotation but in turn the fluctuations produced in \mc\ is about two orders of magnitude larger than that in the 
differential rotation. Indeed helioseismology has detected a significant temporal variation of the \mc\ in last several years 
(e.g., Gonz\'alez-Hern\'andez et al. 2006).
Unfortunately we do not have any measurement of the \mc\ during the \mm. There are some observational studies 
which indicate that solar rotation was different during the \mm\ (Casas et al.\ 2006, and references therein), 
suggesting also the variation of the \mc. 
Some authors (Wang \& Sheeley 2003; 
Passos \& Lopes 2011) suggest weak \mc\ during Maunder minimum. 

Our calculations suggest that the fluctuations in the \mc\ are more important than fluctuations in the 
polar field in inducing grand minima (cf. \S~3.1), although fluctuations in the polar field have
more effect if the dynamo is only moderately critical (cf. \S~3.2). We have seen
that the recovery phase is always gradual and supported by the observation. However, 
depending on the detailed nature of the fluctuations in the \mc\ at the beginning of a grand
minimum, we find that both sudden and gradual initiations of grand minima are possible. Since we
are able to make only a very rough estimate of the coherence time of \mc\ fluctuations, we
explore how our results may change on varying the coherence time (cf. \S~3.5).  For coherence times lying
in the range 20--50 yr, the results remain qualitatively similar.  We also present statistical
analyses of the characteristics of grand minima (cf. \S~3.4). We have seen that both the distributions of the 
waiting times and the durations of the grand minima are exponential, suggesting that these are
governed by the random process. Some of these results are supported by observational 
data (Usoskin et al.\ 2007). 
One issue we did not study here is the north-south asymmetry during grand minima.
There are sufficient evidences that during the \mm\ and the Dalton minimum (Ribes \& Nesme-Ribes 1993; 
Usoskin et al.\ 2009b) there was strong north-south asymmetry in sunspots, indicating this to be a robust feature of grand minima. 
Choudhuri \& Karak (2009) and Karak (2010) have proposed that
if the \pf\ or the \mc\ becomes weak due to the stochastic fluctuations, then it is very unlikely that
they become weak in both the hemisphere by the same amount. With this assumption they have
demonstrated that by introducing a slight asymmetry in the \pf\ or in the \mc\ we can easily model 
the observed north-south asymmetry of sunspots during \mm. We believe that the hemispheric asymmetry in 
grand minima may be another indication for the stochastic forcing as the origin of grand minima.

In this paper we have studied another interesting aspect of solar cycle which is grand maxima (cf. \S~3.6). We have seen 
that similar to grand minima, grand maxima are mostly caused by the strong meridional circulation 
and the distributions of both the waiting times and the durations of the grand maxima are exponential. 
Although the definition of grand maxima is more subjective, we have seen that the average duration 
of grand maxima are shorter compare to that of grand minima. 

One other issue we addressed here is how the Sun comes out of a grand minimum.  The \bl\ mechanism
for the poloidal field generation depends on the existence of sunspots and one naively thinks
that this mechanism would not be operational during the grand minima.  We explored whether the
$\alpha$-effect, which gets suppressed when the toroidal field is strong, could be operational
during the grand minima when the toroidal field becomes weak and whether this $\alpha$-effect
could pull the Sun out of a grand minimum (cf. \S~4). We found that we can match various aspects of observational
data only when we reside this $\alpha$-effect in the upper half of the convection zone and fine-tune its strength. On the other hand, on assuming that 
the \bl\ mechanism remains operational throughout the grand minima, we get results remarkably
close to the observational data.  This raises the provocative question whether the \bl\ 
mechanism could still remain operational during grand minima for reasons we do not understand.
We merely pose this question which cannot be answered at our present level of understanding
of the subject.
 
{\bf Acknowledgements} We thank the referee for careful reading and raising a few comments which improved the manuscript. 
We also thank Dibyendu Nandy and Soumitra Hazra for the discussion which motivated us to study the recovery phase of grand minima presented in Section~4.
ARC thanks DST for partial support through JC Bose Fellowship (project
no.\ SR/S2/JCB-61/2009). BBK thanks Nordita for providing the hospitality where the revised version of the article was prepared.

%%%%%%%%%%%%%%%%%%%%%%%%%%%%%%%%%%%%%%%%%%%%%%%%%%%%%%%%%%%%%%%%%%%%%%%%%%%%%%%%%%%%%%%%
%Effect of quenching on the meridional circulation
%\begin{figure}[!h]
%\includegraphics[width=1.0\textwidth]{}
%\caption{The number of grand minima with a quenching in the meridional circulation.}
%\end{figure}

\label{lastpage}
\end{document}